\documentclass[a4paper, amsfonts, amssymb, amsmath, reprint, showkeys, nofootinbib, twoside, superscriptaddress]{revtex4-1}
\usepackage[english]{babel}
\usepackage[utf8]{inputenc}
\usepackage[colorinlistoftodos, color=green!40, prependcaption]{todonotes}
\usepackage{amsthm}
\usepackage{mathtools}
\usepackage{physics}
\usepackage{xcolor}
\usepackage{graphicx}
\usepackage[left=23mm,right=13mm,top=35mm,columnsep=15pt]{geometry} 
\usepackage{adjustbox}
\usepackage{placeins}
\usepackage[T1]{fontenc}
\usepackage{lipsum}
\usepackage{csquotes}
\usepackage{siunitx}
\usepackage{caption}
\usepackage{subcaption}
\usepackage[pdftex, pdftitle={Article}, pdfauthor={Author}, colorlinks=true]{hyperref}
\begin{document}
\title{Gravitational decoherence by the apparatus in the quantum-gravity induced entanglement of masses}
\author{Fabian Gunnink}
    \affiliation{Van Swinderen Institute for Particle Physics and Gravity, University of Groningen, 9747AG Groningen, the Netherlands }
\author{Anupam  Mazumdar }
    \affiliation{Van Swinderen Institute for Particle Physics and Gravity, University of Groningen, 9747AG Groningen, the Netherlands }
\author{Martine Schut}
    \affiliation{Van Swinderen Institute for Particle Physics and Gravity, University of Groningen, 9747AG Groningen, the Netherlands }
    \affiliation{Bernoulli Institute for Mathematics, Computer Science and Artificial Intelligence, University of Groningen, 9747 AG Groningen, the Netherlands \vspace{1mm}}
\author{Marko Toro\v s }
    \affiliation{School of Physics and Astronomy, University of Glasgow, Glasgow, G12
8QQ, UK}


\begin{abstract}
One of the outstanding questions in modern physics is how to test whether gravity is classical or quantum in a laboratory. Recently there has been a proposal to test the quantum nature of gravity by creating quantum superpositions of two nearby neutral masses, close enough that the quantum nature of gravity can entangle the two quantum systems, but still sufficiently far away that all other known Standard Model interactions remain negligible. However, the mere process of preparing superposition states of a neutral mass (the light system), requires the vicinity of laboratory apparatus (the heavy system). We will suppose that such a heavy system can be modelled as another quantum system; since gravity is universal, the lighter system can get entangled with the heavier system, providing an inherent source of gravitational decoherence. In this paper, we will consider two light and two heavy quantum oscillators, forming pairs of probe-detector systems, and study under what conditions the entanglement between two light systems evades the decoherence induced by the heavy systems. We conclude by estimating the magnitude of the decoherence in the proposed experiment for testing the quantum nature of gravity.
\end{abstract}


\maketitle

\section{Introduction}\label{sec:intro}
The theory of General Relativity (GR) is one of the most well-tested theories of physics, successfully passing a number fundamental tests~\citep{will2014confrontation}, with its latest success being the observation of gravitational waves~\citep{LIGOScientific:2016aoc}. 
However, at short-distance scales and early times, where quantum effects start playing an important role, GR breaks down~\citep{Hawking:1973uf}, and a quantum theory of gravity is needed.
There are several candidate quantum gravity (QG) theories, such as string theory~\citep{Bjerrum-Bohr:2004qcf} and loop quantum gravity~\citep{Thiemann:2006cf}, but despite theoretical progress, the connection with experiments has remained elusive~\citep{amelino2013quantum}.

Albeit the quantization of gravity is an often-used tool in theoretical physics, forming the backbone of candidate quantum-gravity theories, thus far, there is no definitive experimental evidence in support of the quantum nature of gravity.  The reason is simple -- the weakness of the gravitational force makes direct detection of gravitons a formidable challenge, a situation which will likely persist in the foreseeable future~\citep{Dyson:2013hbl}. On the other hand, indirect tests of the quantum nature of gravity (with the first discussions dating back to Feynman~\citep{feynman:1957}) have  in recent years become a real prospect with the advances in precision sensing and metrology, opening the possibility of probing genuine quantum features of gravity with tabletop experiments.

In 2017 a simple experiment for a definitive test of the quantum nature of gravity was proposed in~\citep{Bose:2017nin}, along with its relevant background and feasibility  studies (for a related work see ~\citep{Marletto:2017kzi}). The idea exploits the quantum-gravity-induced entanglement of masses (QGEM) to discern between all classical models of gravity from the quantum one~\footnote{When talking about a theory of quantum gravity, we assume an effective quantum field theory where a massless spin-$2$ graviton acts as a force carrier for the gravitational force, and which behaves well at low energies~\citep{Donoghue:1994dn}.}. Two nearby masses, each delicately prepared in a spatial superposition, are placed close enough that that their mutual gravitational interaction can generate entanglement, but still far enough that all other interactions are strongly suppressed. The generated entanglement can be detected by measuring quantum correlations between the two masses, a genuinely quantum effect with no classical analogue,  and, if detected, would provide the first definite evidence for the quantization of the gravitational field.

The argument for the entanglement-based test of the quantization of gravity can be summarized as follows. To generate matter-matter entanglement one requires a quantum interaction coupling the two systems; the quantum matter-matter gravitational interaction (which in the non-relativistic regime is the operator-valued Newtonian potential) corresponds to the \emph{shift of the energy of the gravitational field}, hence requiring the gravitational field itself to be a quantum operator, ruling out the possibility of a (real-valued) classical gravitational field~\citep{Bose:2022uxe}. Formally, entanglement between two quantum states cannot be increased with local operations and classical communications (LOCC)~\citep{Bennett:1996gf}, as would be the case with a classical gravitational field, and hence, if gravitationally induced entanglement is detected, the gravitational interaction must be ostensibly quantum in nature. This argument has been discussed in detail within the context of perturbative quantum gravity~\citep{Marshman:2019sne,Bose:2022uxe}, the path-integral approach~\citep{Christodoulou2}, and the Arnowitt-Desse-Meissner (ADM) formalism~\citep{Danielson}.  

To discern the spin character of the graviton it is however not sufficient to consider non-relativistic matter-matter interactions but one needs to devise an experiment where gravity couples relativistic fields.
One promising possibility is to probe the quantum light-bending interaction between a heavy mass and photons in a cavity where the degree of the generated entanglement can be used to distinguish between spin $2$ and spin $0$ mediators of the gravitational field~\cite{Biswas:2022qto}.
Another option is to consider matter-matter interactions beyond the static limit where the  post-Newtonian corrections encode the spin character~\citep{Bose:2022uxe}.
   
In this paper we consider the conceptually simple scheme with gravitationally coupled harmonic oscillators and quantify the generated entanglement up to the second post Newtonian contribution.

In order to realise such an experiment one has to overcome are many challenges, such as the preparation of the initial state~\citep{Marshman:2021wyk,Margalit:2020qcy,Zhou:2022frl,Marshman:2021wyk}, the isolation of the system~\citep{vandeKamp:2020rqh,Chevalier:2020uvv,Barker:2022mdz} and the reduction of noise~\citep{Toros:2020dbf}.
The shielding of the system from spurious interactions will never be completely perfect, and the matter systems will loose their coherence due to interaction with the environment. 
Methods for battling the decoherence have been proposed previously~\citep{Schut:2021svd,Tilly:2021qef,Pedrnales}, and many sources of decoherence have been discussed, such as in~\citep{Toros:2020krn,Rijavec:2020qxd,Torrieri:2022znj}.

There is however one source of inherent decoherence which has thus far not been analyzed in detail. In order to witness the generated entanglement we require the presence of nearby experimental apparatus; while electromagnetic couplings between a neutral mass (the light system) and the lab equipment (the heavy system) can be suppressed with appropriate shielding, their mutual gravitational interaction is unavoidable, and scales unfavourably with the mass of the laboratory apparatus. 
The heavy laboratory equipment, which can be modelled quantum mechanically, can entangle with the two neutral masses, thus providing an unavoidable source of gravitational decoherence. 

When we talk about the `apparatus' or `laboratory equipment' we refer to anything close to the experiment that can be quantum, such as the current carrying wires in the Stern-Gerlach setup~\citep{Margalit:2020qcy, Marshman:2021wyk, Zhou:2022jug,Zhou:2022frl}.
We call any such source the `heavy mass', in this paper we consider two heavy systems A and B with mass $m_A=m_B=M$. 
The aim of this paper is to analyze this gravity-induced decoherence in presence of the heavy masses in a model independent fashion, and to quantify the attenuation of the entanglement between the two light quantum masses.

In this paper we will study decoherence with an entanglement measure, the concurrence, which quantifies how much the laboratory equipment and the test masses are entangled.
An often-used approach to analyze decoherence is also to trace out the `environment' system and find the remaining entanglement between the test masses.
We briefly discuss this latter approach in Sec. \ref{sec:results}, but when we talk about `the decoherence' we refer to the entanglement between the apparatus and the test masses. 

First, we will introduce the setup consisting of two heavy quantum harmonic oscillators (representing the laboratory apparatus) and two light quantum harmonic oscillators (representing the two test masses), and introduce all the relevant interactions (Sec.~\ref{sec:setup}).
We will then discuss how to calculate the entanglement using concurrence between the two subsystems that are coupled by the quantized gravitational field within perturbative quantum gravity (Sec.~\ref{sec:ent}). 
Then we discuss the induced decoherence on the two light systems in the static limit (Sec.~\ref{sec:deco}) as well as in the higher order momentum corrections by considering the light systems up to the second post Newtonian contribution (Sec.~\ref{sec:higher_order}). 
We find the allowed parameter space where the entanglement between the light systems dominates the decoherence (Sec.~\ref{sec:results}) and we will conclude with a discussion of the results (Sec.~\ref{sec:conclusion}).

\section{Setup}\label{sec:setup}
Let us consider four massive systems, denoted by $a, b, A, B$ with light masses $m_a,~m_b$ and heavy masses $m_A,~m_B$, respectively. 
We wish to understand the entanglement of $m_a,~m_b$ via the quantum nature of gravity, while $m_A,~m_B$ would be responsible for gravitationally decohering the light masses. 
These massive systems are placed in harmonic traps located at $\pm \frac{d}{2}$ for systems $a, b$ and located at $\pm \frac{D}{2}$ for systems $A,~B$. We will assume $D > d$. 
\begin{figure}[t]
    \centering
    \begin{tikzpicture}[x=0.75pt,y=0.75pt,yscale=-1,xscale=1]
        
        \draw    (157.5,100.5) -- (463.9,100.5);
        \draw   (260.5,137.5) .. controls (260.5,142.17) and (262.83,144.5) .. (267.5,144.5) -- (300.5,144.5) .. controls (307.17,144.5) and (310.5,146.83) .. (310.5,151.5) .. controls (310.5,146.83) and (313.83,144.5) .. (320.5,144.5)(317.5,144.5) -- (353.5,144.5) .. controls (358.17,144.5) and (360.5,142.17) .. (360.5,137.5) ;
        \draw   (462.5,69.5) .. controls (462.5,64.83) and (460.17,62.5) .. (455.5,62.5) -- (320.25,62.5) .. controls (313.58,62.5) and (310.25,60.17) .. (310.25,55.5) .. controls (310.25,60.17) and (306.92,62.5) .. (300.25,62.5)(303.25,62.5) -- (165,62.5) .. controls (160.33,62.5) and (158,64.83) .. (158,69.5) ;
        \draw  [color={rgb, 255:red, 255; green, 255; blue, 255 }  ,draw opacity=1 ][fill={rgb, 255:red, 255; green, 255; blue, 255 }  ,fill opacity=1 ] (200.65,97.85) -- (220.1,97.85) -- (220.1,104.15) -- (200.65,104.15) -- cycle ;
        \draw   (198.68,100.5) -- (205.43,100.5) (221.18,100.5) -- (214.43,100.5) (210.54,97.55) -- (200.32,103.45) (219.54,97.55) -- (209.32,103.45) ;
        \draw  [color={rgb, 255:red, 255; green, 255; blue, 255 }  ,draw opacity=1 ][fill={rgb, 255:red, 255; green, 255; blue, 255 }  ,fill opacity=1 ] (400.9,97.85) -- (420.35,97.85) -- (420.35,104.15) -- (400.9,104.15) -- cycle ;
        \draw   (398.93,100.5) -- (405.68,100.5) (421.43,100.5) -- (414.68,100.5) (410.79,97.55) -- (400.57,103.45) (419.79,97.55) -- (409.57,103.45) ;
        
        \draw (146.5,125.4) node [anchor=south west][inner sep=0.75pt]    {$m_{A}$};
        \filldraw (157.5,100.5) circle (3pt);
        \draw (451,125.4) node [anchor=south west][inner sep=0.75pt]    {$m_{B}$};
        \filldraw (463.5,100.5) circle (3pt);
        \draw (247,125.4) node [anchor=south west][inner sep=0.75pt]    {$m_{a}$};
        \filldraw (259.5,100.5) circle (1.5pt);
        \draw (348.5,125.4) node [anchor=south west][inner sep=0.75pt]    {$m_{b}$};
        \filldraw (361.5,100.5) circle (1.5pt);
        \draw (304,153.4) node [anchor=north west][inner sep=0.75pt]    {$d$};
        \draw (303.5,34.4) node [anchor=north west][inner sep=0.75pt]    {$D$};
    \end{tikzpicture}
    \caption{A graphical representation of the setup that visualizes the introduced parameters $D, d$. With $a$ and $b$ denoting the light systems and $A$ and $B$ denoting the heavy systems.}
    \label{fig:4mass_setup}
\end{figure}
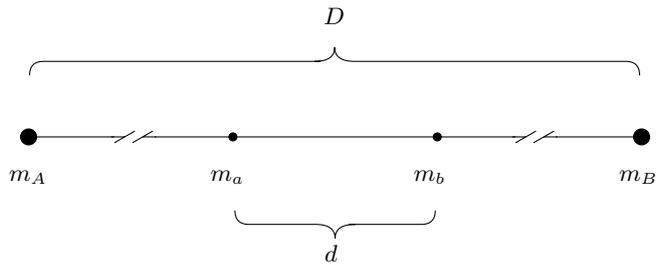
Taking the harmonic oscillators to be well-localized, we obtain:
\begin{align}
    \hat{x}_{a} &= -\frac{d}{2} + \delta\hat{x}_a \, , \qq{} \, \, \,
    \hat{x}_{b} = \frac{d}{2} + \delta\hat{x}_b \, , \label{eq:pos_op_l} \\
    \hat{x}_{A} &= -\frac{D}{2} + \delta\hat{x}_A \, , \qq{}
    \hat{x}_{B} = \frac{D}{2} + \delta\hat{x}_B \, , \label{eq:pos_op_h}
\end{align}
with $\hat{x}_i$ and $\delta \hat{x}_i$ the position operators and small equilibrium displacement for system $i=a,~b,~A,~B$.
We will further assume that all the masses are neutral to minimize the electromagnetic interactions. Although there will be dipolar interactions between all these systems; the Casimir induced dipole-dipole interactions between the two systems $a,~A$ and $b,~B$ can be minimised by placing a conducting plate, while the Casimir interaction between a light and a heavy system can be minimised by 
giving some hierarchy between $D$ and $d$.
The Hamiltonian for the matter systems is given by:
\begin{equation}
    \hat{H}_{\text{m}} = \sum_{i=a,b,A,B} \frac{\hat{p}_i^2}{2m_i} + \frac{m_i \omega_i^2}{2} \delta \hat{x}_i^2 \, ,
\end{equation}
with $\hat{p}_i$ and $\omega_i$ the conjugate momenta and the trap's harmonic frequency for system $i$, respectively.
The basis is chosen such that the matter systems are uncoupled which will simplify our computations~\footnote{As an initial system we can choose a Hamiltonian where there is a coupling between systems $a$ ($b$) and $A$ ($B$):
\begin{align}
    \hat{H}_{\text{m}} = \sum_{i=1,2,3,4} \frac{\hat{p}_i'^2}{2m_i} &+ \frac{k_0}{2} (\delta \hat{x}_1^2 + \delta \hat{x}_2^2) + \frac{k_1}{2} (\delta \hat{x}_1 - \delta \hat{x}_2)^2 \nonumber \\ &+ \frac{k_2}{2} (\delta \hat{x}_3^2 + \delta \hat{x}_4^2) + \frac{k_3}{2} (\delta \hat{x}_3 - \delta \hat{x}_4)^2 \, .
\end{align}
Then there exists a unitary transformation such that the Hamiltonian becomes decoupled.
After the transformation the matter Hamiltonian can be written as 
\begin{equation}
    \hat{H}_{\text{m}} = \hat{H}_a + \hat{H}_b + \hat{H}_A + \hat{H}_B \, . 
\end{equation}
with $\hat{H}_i = \frac{\hat{p}_i^2}{2m_i} + \frac{1}{2} m_i\omega_i^2\hat{x}_i^2$, and with $\omega_{a(b)}^2 = k_{0(2)}/m_{a(b)}$, $\omega_{A(B)}^2 = [k_{0(2)} + 2 k_{1(3)}]/m_{A(B)}$.
The change of basis is given as:
$$\hat{x}_{a(b)} = [\hat{x}_{1(3)} + \hat{x}_{2(4)}]/\sqrt{2}, \qq{} \hat{x}_{A(B)} = [\hat{x}_{1(3)} - \hat{x}_{2(4)}]/\sqrt{2} \, .$$
}.
The mode operators for the harmonic oscillator systems are given by:
\begin{align}
    \delta\hat{x}_j = \sqrt{\frac{\hbar}{2 m_j \omega_j}}(j + j^\dagger) \, , \, \,
    \hat{p}_j = i \sqrt{\frac{\hbar m_j \omega_j}{2}}(j - j^\dagger) \, , \label{eq:mode_op}
\end{align}
with $j=a,b,A,B$, and the operators satisfying the usual commutation relations~\footnote{
These commutation relations are:
\begin{align*}
    &[a,a] = [b,b] = [A,A] = [B,B] = 0 \\
    &[a^\dagger, a^\dagger] = [b^\dagger, b^\dagger] = [A^\dagger, A^\dagger] =[B^\dagger, B^\dagger] = 0 \\
    &[a, a^\dagger] = [b, b^\dagger] = [A, A^\dagger] = [B, B^\dagger] = 1 \, .
\end{align*}
}.
Thus the Hamiltonian can be written as:
\begin{equation}
    \hat{H}_{\text{m}} = \hbar \omega_a \hat{a}^\dagger \hat{a} + \hbar \omega_b \hat{b}^\dagger \hat{b} + \hbar \omega_A \hat{A}^\dagger \hat{A} + \hbar \omega_B \hat{B}^\dagger \hat{B} \, .
\end{equation}
We now introduce a gravitational field and study the interaction Hamiltonian $\hat{H}_\text{int}$ between the gravitational and matter fields. 

We work in linearized gravity where the metric is given by $g_{\mu\nu} = \eta_{\mu\nu} + h_{\mu\nu}$, with $\eta_{\mu\nu}$ the flat Minkowski background with signature $(-,+,+,+)$ and with $h_{\mu\nu}$ a perturbation which is small in magnitude around the Minkowski background.
The metric fluctuations are then promoted to quantum operators:
\begin{align}
    \hat{h}_{\mu\nu} = \mathcal{A} \int \dd[3]{k} \sqrt{\frac{\hbar}{2\omega_k (2\pi)^3}} \left( \hat{P}_{\mu\nu}^\dagger(\vec{k})e^{-i\vec{k}\vec{r}} + \text{H.c.} \right) ,
\end{align}
with $\mathcal{A}= \sqrt{16 \pi G/c^2}$, and where $\hat{P}_{\mu\nu}$ and $\hat{P}_{\mu\nu}^\dagger$ denote the graviton annihilation and creation operators, respectively, and satisfy the following commutation relations~\citep{Gupta-1952}:
\begin{align}\label{eq:grav_comm_rel}
    [\hat{P}_{\mu\nu}(\vec{k}), \hat{P}_{\rho\sigma}^\dagger(\vec{k}')] = (\eta_{\mu\rho} \eta_{\nu\sigma} + \eta_{\mu\sigma} \eta_{\nu\rho}) \delta(\vec{k} -\vec{k}') \, .
\end{align}
In the weak field regime we can decompose the metric fluctuation operator into two modes: the spin-$2$ mode $\gamma_{\mu\nu}$ and the spin-0 mode $\gamma \equiv \eta_{\mu\nu} \gamma^{\mu\nu}$~\citep{Gupta-1952}~\footnote{
These two modes can be treated independently. $\gamma_{\mu\nu}$ is sometimes called the trace-reversed metric since $h=-\gamma$.
}.
Such that:
$
    \hat{h}_{\mu\nu} = \hat{\gamma}_{\mu\nu} - \frac{1}{2} \eta_{\mu\nu} \hat{\gamma} \, .
$
Consequently the spin-$2$ and spin-$0$ decomposed parts of the graviton can be promoted to operators as well, and they are given in terms of the graviton creation- and annihilation operators~\citep{Gupta-1952}:
\begin{align}
    \hat{\gamma}_{\mu\nu} = \mathcal{A} \int \dd[3]{k} \sqrt{\frac{\hbar}{2\omega_k (2\pi)^3}} \left( \hat{P}_{\mu\nu}^\dagger(\vec{k})e^{-i\vec{k}\vec{r}} + \text{H.c.} \right) , \\
    \hat{\gamma} = 2 \mathcal{A} \int \dd[3]{k} \sqrt{\frac{\hbar}{2\omega_k (2\pi)^3}} \left( \hat{P}^\dagger(\vec{k})e^{-i\vec{k}\vec{r}} + \text{H.c.} \right) ,
\end{align}
satisfying the commutation relations in Eq. \eqref{eq:grav_comm_rel}~\footnote{
Following Eq. \eqref{eq:grav_comm_rel} and the definition $\gamma \equiv \eta_{\mu\nu}\gamma^{\mu\nu}$, the additional commutation relation is:
\begin{align}
    [\hat{P}(\vec{k}), \hat{P}^\dagger(\vec{k}')] = - \delta(\vec{k} -\vec{k}') \, .
\end{align}
}.
The gravity Hamiltonian can then be written in terms of graviton creation and annihilation operators~\citep{Gupta-1952}.
%
Now that both the matter and graviton systems have been introduced, we continue by studying their interaction and in the next section the consequential entanglement generation.
The interaction term is given by the graviton coupling to the stress-energy tensor $\hat{T}_{\mu\nu}$ (which specifies the matter system contents):
\begin{equation}
    \hat{H}_{\text{int}} = - \frac{1}{2} \int \dd[3]{r} \hat{h}^{\mu\nu}(\vec{r}) \hat{T}_{\mu\nu}(\vec{r}) \, .
\end{equation}
We consider the two harmonically trapped particles $a, b$ to be moving along the $x$-axis, and the two heavy systems $A, B$ to be static.
The systems $A, B$ are taken to be static because we consider these systems to be the very massive systems such that their motion remains negligible when perturbed by the two light systems.
The four systems thus generate the following currents:
\begin{align}
    \hat{T}_{00}(\vec{r}) \equiv \sum_{n=a,b,A,B} m_n c^2 \delta( \vec{r} - \hat{\vec{r}}_n) \, , \label{eq:T00_pos} \\
    \hat{T}_{ij}(\vec{r}) \equiv \sum_{n=a,b} \frac{\hat{p}_{n,i}\hat{p}_{n,j}}{E/c^2} \delta( \vec{r} - \hat{\vec{r}}_n) \label{eq:Tij_pos}\, ,
\end{align}
with the position of the matter systems $\hat{r}_n = (\hat{x}_n, 0, 0)$, with the momentum $\hat{p}_{\mu} = (-E/c, \vec{p})$, energy $E = \sqrt{\vec{p}^2c^2 + m^2 c^4}$, and with $i,~j=1,2,3$. 

Since we specified the movement of the oscillators $a, b$ to be along the $x$-axis, the only non-zero $\hat{T}_{\mu\nu}$-components are $\hat{T}_{01}$, $\hat{T}_{10}$ and $\hat{T}_{11}$.
Therefore, the only relevant $\hat{h}_{\mu\nu}$ components in the coupling are $\hat{h}_{00} = \hat{\gamma}_{00} + \frac{1}{2}\hat{\gamma}$, $\hat{h}_{01} = \hat{h}_{10} = \hat{\gamma}_{01}$ and $\hat{h}_{11} = \hat{\gamma}_{11} - \frac{1}{2}\hat{\gamma}$.
Writing the interaction Hamiltonian in terms of the decomposed metric perturbation, while exploiting the symmetries $\hat{T}_{01} = \hat{T}_{10}$ and $\hat{\gamma}_{01} = \hat{\gamma}_{10}$, gives:
\begin{align}\label{eq:H_int}
    \hat{H}_{\text{int}} &=\int\dd[3]{r} \bigg( \frac{1}{2} [ \hat{\gamma}_{00}(\vec{r}) + \frac{1}{2}\hat{\gamma}(\vec{r})] \,\hat{T}_{00}(\vec{r}) \nonumber \\
    &+ \frac{1}{2} [ \hat{\gamma}_{11}(\vec{r}) - \frac{1}{2}\hat{\gamma}(\vec{r})]\, \hat{T}_{11}(\vec{r}) + \hat{\gamma}_{10}(\vec{r}) \, \hat{T}_{10}(\vec{r}) \bigg).
\end{align}
As explained in Ref.~\citep{Bose:2022uxe}, the energy shift in the graviton vacuum due to the above interaction
can only induce entanglement when the gravitational field is quantized, with $h_{\mu\nu}$ or equivalently $\gamma_{\mu\nu},~\gamma$.
This can be formalized using the Local Operations and Classical Communication (LOCC) principle, which states that a LOCC channel (such is the case for a classical real valued gravitational field) cannot increase the entanglement between the two systems.
Only Quantum Communication can increase entanglement between the systems~\citep{Bose:2022uxe}.
The graviton here acts as a quantum communicator between the two systems, and is therefore able to induce a coupling that entangles previously unentangled oscillators.
This entanglement and decoherence are studied in the next sections.

\section{Entanglement via graviton}\label{sec:ent}
We assume that initially the quantum matter systems are in the ground state (denoted by $\ket{0}_i$, with $i$ specifying the system $i=a,~b,~A,~B$):
\begin{equation}
    \ket{\psi_i} = \ket{0}_a \ket{0}_b \ket{0}_A \ket{0}_B \, .
\end{equation}
Since gravity will couple all the systems, it will induce interaction between the heavy and light oscillators, $\hat H_{hl}$ (which is presented in Eqs. \eqref{eq:ham_1order} and \eqref{eq:next_order_Hhl}).
As a result of this interaction the final state will evolve to:~\footnote{
Here we have left out the subscripts on the kets to ease the notation.
In the remainder of the paper the order of the the states is always $a,~b,~A,~B$.
}
\begin{equation}
    \ket{\psi_f} = \frac{1}{\sqrt{\mathcal{N}}} \sum_{\substack{n_a, n_b \\ n_A, n_B}} C_{n_a n_b n_A n_B} \ket{n_a} \ket{n_b} \ket{n_A} \ket{n_B} \, . \label{eq:psi_f}
\end{equation}
The number states are denoted by $\ket{n_i}$, and the normalisation is given by $\mathcal{N} = \sum_{n_a, n_b \\ n_A, n_B} \abs{C_{n_a n_b n_A n_B}}^2$.
The interaction is scaled by a bookkeeping parameter $\lambda $.

In first order perturbation theory the coefficients for the final wavefunction are given by:
\begin{equation}\label{eq:coefficients}
    C_{n_a n_b n_A n_B} = \lambda \frac{\bra{n_a}\bra{n_b}\bra{n_A}\bra{n_B}\hat{H}_{hl}\ket{0}\ket{0}\ket{0}\ket{0}}{\sum_{i=a,b,A,B} (E_{0_i} - E_{n_i})}\, ,
\end{equation}
for the perturbed states, and $C_{0000}=1$ for the unperturbed state.
In the above equation $E_{0_i}$ is the ground-state energy and $E_{n_i}$ denotes the $n$th excited state energy, for system $i=a,b,A,B$.  

At this point it is important to take a note that $\hat{H}_{hl}$ is a quantum operator. If it were classical, so not operator-valued, then for any perturbed coefficients $C_{n_a n_b n_A n_B} = 0$ due to the orthogonality of the states, thus the final wavefunction would be $\ket{\psi_f} = \ket{0} \ket{0} \ket{0} \ket{0}$, the initial wavefunction. No entanglement can be generated in an initially unentangled system from a classical interaction.
Since we are working in the framework of perturbative quantum field theory of gravity we expect an entanglement, which will be quantified by the concurrence of a biparte system, between the subsystems $1$ and $2$ (see below for the choice of the subsystems 1 and 2):
\begin{equation}\label{eq:conc_def}
    \mathcal{C} \equiv \sqrt{2-2\Tr(\rho_1^2)} \, ,
\end{equation}
where $\rho_1 = \Tr_2(\rho)$ is the partial density matrix found by tracing out subsystem $2$ in the full density matrix $\rho = \ket{\psi_f}\bra{\psi_f}$.
The larger the concurrence, the more strongly entangled the subsystems are, where a maximally entangled state gives the value $\sqrt{2}$ and an unentangled state gives the value $0$~\footnote{The concurrence can be related to the maybe better-known and more widely applicable von Neumann entropy via a simple relation~\citep{Wootters:1997id}.}.

The use of concurrence is limited to biparte systems though.
As we are interested in the decoherence of the systems $a,~b$ due to their coupling to the more massive systems $A,~B$, we choose the bipartition such that subsystem $1$ consists of the light oscillators $a,~b$ and subsystem $2$ consists of the heavy oscillators $A,~B$.
Since the entanglement and decoherence are two sides of the same coin, by studying the concurrence for this bipartition we gain information about the effects of the apparatus (the heavy oscillators) on the coherence of the QGEM experiment (the two light particles).
For the light-heavy bipartition, the partial density matrix for the light system is
\begin{equation}\label{eq:density_matrix_ll}
    \rho_1 = \frac{1}{\mathcal{N}} \sum_{\substack{n_a, n_b, N_a, \\ N_b, n_A, n_B}} C_{n_a n_b n_A n_B} C^*_{N_a N_b n_A n_B} \ket{n_a \, n_b}\bra{N_a \, N_b}
\end{equation}
using the notation $\ket{n_a \, n_b} = \ket{n_a} \ket{n_b}$.
Inserting this expression into Eq. \eqref{eq:conc_def}, the heavy-light concurrence, denoted $\mathcal{C}_\text{hl}$, can be expressed in terms of the coefficients $C$ defined in Eq. \eqref{eq:coefficients}:
\begin{align}\label{eq:concurrence_def_pure}
    \mathcal{C}_\text{hl} \equiv \bigg[ 2-\frac{2}{\mathcal{N}^2} &\sum_{\substack{n_a, N_a, n_b, N_b \\ n_A, n_B, N_A, N_B}} C_{n_a n_b n_A n_B} C^*_{N_a N_b n_A n_B} \nonumber \\ &\hspace{7mm} \cross C_{n_a n_b N_A N_B} C^*_{N_a N_b N_A N_B} \bigg]^{1/2}.
\end{align}
Finding all the relevant expressions of the coefficients in Eq. \eqref{eq:coefficients} would result in the quantification of decoherence/entanglement at first order in the perturbation theory. 
For this we need to find the interaction Hamiltonian between the heavy and light system, $\hat{H}_\text{hl}$, which is generated by the exchange of the virtual graviton (see below the derivation with the result in Eq. \eqref{eq:ham_1order}).

The interaction between gravity and matter is given in Eq. \eqref{eq:H_int}, from which we can compute the shift in energy to the graviton vacuum at second order in perturbation theory~\footnote{
The first order term  corresponding to the emission/absorption of a graviton is given by $\bra{0}\hat{H}_\text{int}\ket{0}$.
This contribution vanishes since $\hat{H}_\text{int}$ depends linearly on the graviton creation and annihilation operators, and $\hat{P}\ket{0}=\hat{P}_{\mu\nu}\ket{0}=0$, $\bra{0}\hat{P}^\dagger = \bra{0}\hat{P}_{\mu\nu}^\dagger= 0$. 
In the second order term (corresponding to the exchange of a virtual graviton) $\bra{0}\hat{H}_\text{int}|\vec{k}\rangle$ is quadratically dependent on the creation an annihilation operators.
Using the operator commutation rules shows that this contribution is non-vanishing.
}:
\begin{equation}
    \Delta \hat{H}_g \equiv \int \dd[3]{k} \frac{\bra{0}\hat{H}_\text{int}|\vec{k}\rangle\langle\vec{k}|\hat{H}_\text{int}\ket{0}}{E_0 - E_k} \, ,\label{eq:DeltaHg}
\end{equation}
with $E_0$ the energy of the vacuum state, and $E_k = E_0 + \hbar \omega_k$ the energy of of the one-particle state $|\vec{k}\rangle$ representing the intermediate graviton, which is created from the vacuum with the graviton creation operators.
The collection of normalized projectors $|\vec{k}\rangle \langle\vec{k}|$ is given by:
\begin{align}
    |\vec{k}\rangle\langle\vec{k}| &= 
    \frac{1}{2} P^\dagger_{00}(\vec{k})\ket{0}\bra{0}P_{00}(\vec{k}) + \frac{1}{2} P^\dagger_{11}(\vec{k})\ket{0}\bra{0}P_{11}(\vec{k}) \nonumber\\ &- P^\dagger_{01}(\vec{k})\ket{0}\bra{0}P_{01}(\vec{k}) - P^\dagger(\vec{k})\ket{0}\bra{0}P(\vec{k}).
\end{align}
For each projector summed in the above expression we can evaluate $\bra{0}\hat{H}_\text{int}|\vec{k}\rangle$, with the interaction given in Eq. \eqref{eq:H_int}:
\begin{align}
    \bra{0}\hat{H}_\text{int}\hat{P}_{00}(\vec{k})|\vec{0}\rangle &= \mathcal{A} \sqrt{\frac{\hbar }{2\omega_k}} \hat{T}_{00}(\vec{k}) \, , \label{eq:HP00}\\
    \bra{0}\hat{H}_\text{int}\hat{P}_{11}(\vec{k})|\vec{0}\rangle &= \mathcal{A} \sqrt{\frac{\hbar }{2\omega_k}} \hat{T}_{11}(\vec{k}) \, , \\
    \bra{0}\hat{H}_\text{int}\hat{P}_{01}(\vec{k})|\vec{0}\rangle &= \mathcal{A} \sqrt{\frac{\hbar }{2\omega_k}}\hat{T}_{01}(\vec{k}) \, , \\
    \bra{0}\hat{H}_\text{int}\hat{P}(\vec{k})|\vec{0}\rangle &= \frac{\mathcal{A}}{2} \sqrt{\frac{\hbar }{2\omega_k}}\left[\hat{T}_{00}(\vec{k}) - \hat{T}_{11}(\vec{k})\right] \, , \label{eq:HP}
\end{align}
with $\mathcal{A}\equiv\sqrt{16\pi G/c^2}$. 
$\hat{T}_{\mu\nu}(\vec{k})$ are the stress-energy tensor components in momentum space~\footnote{
The momentum-space stress-energy tensor components are given by the Fourier transform of the components in position space:
\begin{equation}
    \hat{T}_{\mu\nu}(\vec{k}) = \frac{1}{(2\pi)^{3/2}} \int\dd[3]
    {r} e^{-i \vec{k}\cdot \vec{r}} \hat{T}_{\mu\nu}(\vec{r})\, .
\end{equation}
}
, which from Eqs. \eqref{eq:T00_pos},\eqref{eq:Tij_pos} are found to be:
\begin{align}
    \hat{T}_{00} (\Vec{k}) &= \frac{1}{(2\pi)^{3/2}} \bigg[ m_{A}c^2  e^{-i\Vec{k} \cdot \hat{r}_A} + m_{B}c^2 e^{-i\Vec{k} \cdot \hat{r}_B} \nonumber \\ 
    &\qq{}\qq{}\qq{}+ E_{a} e^{-i\Vec{k} \cdot \hat{r}_a} + E_{b} e^{-i\Vec{k} \cdot \hat{r}_b} \bigg], \label{eq:T00}\\[.5em]
    \hat{T}_{01} (\Vec{k}) &= - \frac{c}{(2\pi)^{3/2}} \left[ \hat{p}_{a} e^{-i\Vec{k} \cdot \hat{r}_a} + \hat{p}_{b} e^{-i\Vec{k} \cdot \hat{r}_b} \right], \\[.5em]
    \hat{T}_{11} (\Vec{k}) &= \frac{1}{(2\pi)^{3/2}} \left[ \frac{\hat{p}_{a}^2 c^2}{E_a} e^{-i\Vec{k} \cdot \hat{r}_a} + \frac{\hat{p}_{b}^2 c^2}{E_b} e^{-i\Vec{k} \cdot \hat{r}_b} \right]\label{eq:T11}.
\end{align}
Filling in Eqs. \eqref{eq:T00}-\eqref{eq:T11} and Eqs. \eqref{eq:HP00}-\eqref{eq:HP} into Eq. \eqref{eq:DeltaHg} gives an expression for the graviton energy shift from the vacuum, $\Delta \hat{H}_g$.
This expression can be simplified by performing the integral over $\vec{k}$~\footnote{
This integration is simply
\begin{equation}
    \int \frac{\dd[3]{k}}{(2\pi)^3} \frac{1}{|\vec{k}|^2} e^{i\vec{k}\cdot\hat{\vec{r}}} = \frac{1}{4 \pi \hat{\vec{r}}} \, ,
\end{equation}
and the expression was rewritten such that $\hat{\vec{r}} = \hat{\vec{x}}_i - \hat{\vec{x}}_j$.
}. 
Furthermore we restrict the movement to the $x$-axis, meaning that $\hat{p}_{i,y} = \hat{p}_{i,z} = 0$, $\hat{p}_{i,x} \equiv \hat{p}_{i}$ and $\hat{r}_i=(\hat{x}_i,0,0)$ for $i=a,b,A,B$, to find the expression:
\begin{align}
    \Delta \hat{H}_g &= - \frac{\mathcal{A}^2}{16 \pi c^2} \Bigg[ \frac{ m_{A} E_{a} c^2 + m_{A} \frac{\hat{p}_{a}^2 c^4}{E_{a}}}{\abs{\hat{x}_A - \hat{x}_a}} \nonumber \\ 
    &+  \frac{m_{A} E_{b} c^2 + m_{A} \frac{\hat{p}_{b}^2 c^4}{E_{b}}}{\abs{\hat{x}_{A} - \hat{x}_{b}}}
    + \frac{m_{A}m_{B}c^4}{\abs{\hat{x}_{A} - \hat{x}_{B}}} \nonumber \\[.5em]
    &+ \frac{E_{a}E_{b} - 4 \hat{p}_{a}\hat{p}_{b}c^2 + \hat{p}_{a}^2 c^2\frac{E_{b}}{E_{a}} + \hat{p}_{b}^2 c^2\frac{E_{a}}{E_{b}} + \frac{\hat{p}_{a}^2 \hat{p}_{b}^2 c^4}{E_{a}E_{b}}}{\abs{\hat{x}_{a} - \hat{x}_{b}}}  \nonumber \\[.5em]
    &+ \frac{ m_{B}E_{a}c^2 + m_{B} \frac{\hat{p}_{a}^2 c^4}{E_{a}}}{\abs{ \hat{x}_{a} - \hat{x}_{B}}} + \frac{ m_{B}E_{b}c^2 + m_{B} \frac{\hat{p}_{b}^2 c^4}{E_{b}} }{\abs{ \hat{x}_{b} - \hat{x}_{B}}} \Bigg] .\label{eq:H_g}
\end{align}
Taking $m_a = m_b = m$ and $m_A = m_B = M$, and expanding Eq. \eqref{eq:H_g} in powers of $1/c^2$ gives the non-relativistic couplings among the $4$ oscillators upto order $1/c^4$, and in first order in $G$, the full expression is presented in Eq. \eqref{eq:H_g_op}
~\footnote{
We can reach the  \textit{classical} point particle limit by substituting $\vec{r}\equiv \vec{x_i}-\vec{x_j}$ with the number-valued distances discussed in Sec. \ref{sec:setup}, the potential becomes:
\begin{align}
    \Delta H_g 
    &= - G \bigg[ \frac{m^2}{d} + \frac{M^2}{D} - \frac{8 m M}{d^2-D^2} \bigg] \nonumber \\
    &- \frac{G}{c^2} \bigg[ \frac{3p_a^2 - 8p_ap_b + 3p_b^2}{2d} - \frac{6 D M (p_a^2+p_b^2)}{(d^2-D^2) m} \bigg] \nonumber \\
    &- \frac{G}{c^4} \bigg[\frac{5p_a^4-18p_a^2p_b^2 +5p_b^4}{8dm^2} 
    - \frac{20 D M(p_a^4+p_b^4)}{8(d^2-D^2)m^3}\bigg] \nonumber \\ &+\order{\frac{1}{c^6}}\, . \label{eq:H_g_NR}
\end{align}
If the heavy systems are not taken into account, i.e. $M=0$, Eq. \eqref{eq:H_g_NR} reduces to the same expression found in Ref.~\citep{Bose:2022uxe} for the interaction between two harmonic oscillators.
Furthermore, in the center-of-mass frame, i.e. $p\equiv p_a = -p_b$, Eq. \eqref{eq:H_g_NR} gives a potential that matches known results for the non-relativistic potential between classical point particles~\citep{Grignani:2020ahv,Iwasaki:1971vb,Cristofoli:2019neg}.}.

\section{Quantifying the decoherence}\label{sec:deco}
In this section we give the expression for the decoherence due to the gravitational interaction between the heavy and light systems. 
We find the decoherence using an entanglement measure, the concurrence, given in Eq. \eqref{eq:concurrence_def_pure}, which quantifies the information of the light system shared with the heavy system.
We start by finding the first order interaction terms between the heavy and light systems.
We can substitute the expressions \eqref{eq:pos_op_l},\eqref{eq:pos_op_h} for the position operators in terms of their displacements
into the Hamiltonian in Eq. \eqref{eq:H_g_op}, and look at the lowest order coupling between the light and heavy matter systems~\footnote{
Since we are considering a bipartite heavy-light system, only the interaction between heavy and light is taken into account to find the decoherence.
Any heavy-heavy or light-light interaction can be viewed as `self-interaction' since it only causes entanglement within the subsystem.
However, the strength of the light-light entanglement is important to analyse the decoherence effects of the heavy system.
Taking into account only the light-light couplings in Eq. \eqref{eq:H_g_op} and following the same procedure as described in this section, we find the concurrence between the two light oscillators at lowest order to be:
\begin{align}
    \mathcal{C}_\text{ll} &= \frac{G m}{d^3 \omega_l^2} + \frac{2G m}{c^2 d}  \, , \label{eq:lconc1} 
\end{align}
where we have taken the first order coupling, which consists of a static contribution (from the position operator coupling) and a non-static contribution (from momentum operator coupling), with the momentum contribution being suppressed by $1/c^2$.
\label{fn:bipartite_condits}
}.
These can be found by Taylor expanding the small displacements $\delta \hat{x}_i$, giving the lowest order interaction terms:
\begin{equation}\label{eq:ham_1order}
    \hat{H}_{hl} = 16 G m M \bigg[ \frac{\delta\hat{x}_a\delta\hat{x}_A+\delta\hat{x}_b\delta\hat{x}_B}{(D-d)^3} + \frac{\delta\hat{x}_A\delta\hat{x}_b + \delta\hat{x}_a\delta\hat{x}_B}{(D+d)^3} \bigg]
\end{equation}
Note that in the above expression there is no coupling between the momentum and the position operators, even though the light system is taken to be non-static.
This is because the lowest order coupling is between \textit{one} heavy position/momentum operator and \textit{one} light position/momentum operator. 
The coupling with momentum operators at this order appears as $- 4 G \hat{p}_a \hat{p}_b / d c^2$, it only gives a coupling between the two light particles instead of the light and heavy subsystems.

We will now use the mode operators in Eq. \eqref{eq:mode_op} to write $\hat{H}_{hl}$ in terms of the mode operators $j,j^\dagger$ with $j=a,b,A,B$.
The resulting Hamiltonian is:
\begin{align} \label{eq:ham_mode_op}
\hat{H}_{hl}^{\text{op}} = 
    \frac{8 G \hbar \sqrt{M m}}{\sqrt{\omega_l \omega_h}}
    \left[ \frac{a^\dagger A^\dagger + b^\dagger B^\dagger}{(D-d)^3} + \frac{a^\dagger B^\dagger + A^\dagger b^\dagger}{(D+d)^3} \right] \, ,
\end{align}
where all irrelevant terms (the terms that annihilate the vacuum) have been left out for simplicity. 
Filling the Hamiltonian in Eq. \eqref{eq:ham_mode_op} into Eq. \eqref{eq:coefficients}, we find the only non-zero coefficients are~\footnote{
Any terms of the form $\ket{0\, 0 \, n_A \, n_B}$ and $\ket{n_a \, n_b\,0\,0}$ are omitted because they arise from the self-interaction within the light and heavy subsystems, respectively, and are therefore not relevant to our analysis. 
\label{fn:coef_condits}
} 
(where we assumed $\omega_a = \omega_b = \omega_l$ and $\omega_A = \omega_B = \omega_h$ for simplicity, and set $\lambda$ = 1):
\begin{align}
    C_{1010} &= C_{0101} = -\frac{\mathfrak{g}_{-}}{\omega_h + \omega_l}\, , \\ 
    C_{0110} &= C_{1001}= -\frac{\mathfrak{g}_{+}}{\omega_h + \omega_l} 
\end{align}
with 
\begin{align}
    \mathfrak{g}_{\pm} &= \frac{8 G}{(D\pm d)^3} \frac{\sqrt{m M}}{\sqrt{\omega_h \omega_l}} \, .
    \label{eq:couplings}
\end{align}
The final state given in Eq.~\eqref{eq:psi_f} (up to the first order in the perturbation theory) is thus given by:
\begin{align}
    \ket{\psi_f} = &\frac{1}{\sqrt{\mathcal{N}}} \Big( \ket{0000} -\frac{\mathfrak{g}_{-}}{\omega_h + \omega_l} \ket{1010} - \frac{\mathfrak{g}_{+}}{\omega_h + \omega_l} \ket{0110} \nonumber \\ &-\frac{\mathfrak{g}_{+}}{\omega_h + \omega_l} \ket{1001} -\frac{\mathfrak{g}_{-}}{\omega_h + \omega_l} \ket{0101} \Big)\, , \label{eq:psi_f_full}
\end{align}
with the normalization $\mathcal{N}= 1 + 2 (\mathfrak{g}_{-}^2 + \mathfrak{g}_{+}^2)/(\omega_h+\omega_l)^2$, and using the notation $\ket{n_a}\ket{n_b}\ket{n_A}\ket{n_B} = \ket{n_a\,n_b\,n_A\,n_B}$.
The final state is an entangled state between the ground states and first excited states of the light and heavy subsystems.
Due to the pair-wise interactions taken here, in each of the perturbed states one heavy and one light system are in the first excited states.
Using Eq.~\eqref{eq:conc_def} the concurrence is found to be:
\begin{equation}
    \mathcal{C}_\text{hl}^{(1)} = \sqrt{2-2\frac{1+2\frac{(\mathfrak{g}_-^2 + \mathfrak{g}_+^2)^2}{\omega^4} + 8\frac{\mathfrak{g}_-^2\mathfrak{g}_+^2}{\omega^4}}{\big(1+2 \frac{\mathfrak{g}_-^2 + \mathfrak{g}_+^2}{\omega^2} \big)^2}} \, , \label{eq:conc_two_op}
\end{equation}
where $\omega \equiv \omega_h +\omega_l$ for simplicity, and the superscript $(1)$ denotes that we have taken the lowest order contributions to the entanglement (i.e., linear equations of motion).
In the limit where $\mathfrak{g}_\pm / \omega \ll 1$ the concurrence becomes
\begin{equation}\label{eq:conc_first_order}
    \mathcal{C}_\text{hl}^{(1)} \approx 2 \sqrt{2} \sqrt{\frac{\mathfrak{g}_+^2+\mathfrak{g}_{-}^2}{\omega^2}} \, .
\end{equation}
We now consider two special cases representing different experimental setups: $D\gg d$ and $D=2d$. 
Taking the limit $D \gg d$, we can Taylor expand the couplings $\mathfrak{g}_\pm \approx \frac{8 G}{D^3} \frac{\sqrt{m M}}{\sqrt{\omega_h \omega_l}}\left( 1 \mp 3 \frac{d}{D} + \order{\frac{d^2}{D^2}}\right)$.
The expression for the concurrence simplifies to:~\footnote{It turns out that when keeping these second order terms, the approximation $\mathfrak{g}_\pm/\omega \ll 1$ simplifies the concurrence to:
$$ \mathcal{C}_\text{hl}^{(1)}(D\gg d) \approx 2 \sqrt{2} \frac{\mathfrak{g}}{\omega} \sqrt{2 + 42 \frac{d^2}{D^2}} \approx 4 \frac{\mathfrak{g}}{\omega} \left( 1 + \frac{21}{2} \frac{d^2}{D^2} \right) \, .$$
Neglecting $\order{(d/D)^2}$ terms we recover eq. \eqref{eq:concD}.
}
\begin{equation}\label{eq:concD}
    \mathcal{C}_\text{hl}^{(1)}(D\gg d) \approx \frac{32 G}{(\omega_h+\omega_l)D^3} \sqrt{\frac{Mm}{\omega_h\omega_l}}\, .
\end{equation}
The degree of entanglement grows with the masses of the light and heavy system ($m,~M$, respectively), but it grows inversely with the harmonic trap frequencies and inversely (inverse cubic) with the distance between the light and heavy system. 

We now explore another possible configuration of the four oscillators where the spacing between any adjacent oscillators will be $d$, by setting $D=2d$.
In this case the concurrence in Eq.~\eqref{eq:conc_first_order} simplifies to:
\begin{align}\label{eq:concd}
    \mathcal{C}_\text{hl}^{(1)}(D=2d) \approx \frac{32 \sqrt{365} G}{27 (\omega_h+\omega_l) d^3} \sqrt{\frac{Mm}{\omega_h\omega_l}} \, .
\end{align}
We note that the dependence on the masses, frequencies and distance between the oscillators is identical to the behaviour of the concurrence in Eq.~\eqref{eq:concD}.


Instead of limiting $D$ and choosing a specific setup to simplify the results, we could also note that the coupling between two neighbouring oscillators, $\mathfrak{g}_-$, will dominate over the coupling between two maximally separated oscillators, $\mathfrak{g}_+$, (see the denominator of the first factor in Eq. \eqref{eq:couplings}).
We can thus simplify the expression of the concurrence by considering only the coupling $\mathfrak{g}=\sqrt{\mathfrak{g}_+^2+\mathfrak{g}_-^2} \approx\mathfrak{g}_{-}$ (and $\mathfrak{g}_-$ as in Eq. \eqref{eq:couplings}), giving:
\begin{equation}\label{eq:concg}
    \mathcal{C}_\text{hl}^{(1)}(\mathfrak{g}) \approx \frac{16\sqrt{2}G}{(D-d)^3(\omega_h+\omega_l)} \sqrt{\frac{2 m M}{\omega_h \omega_l}} \, .
\end{equation}

\begin{figure}[t]
     \centering
        \includegraphics[width=\linewidth]{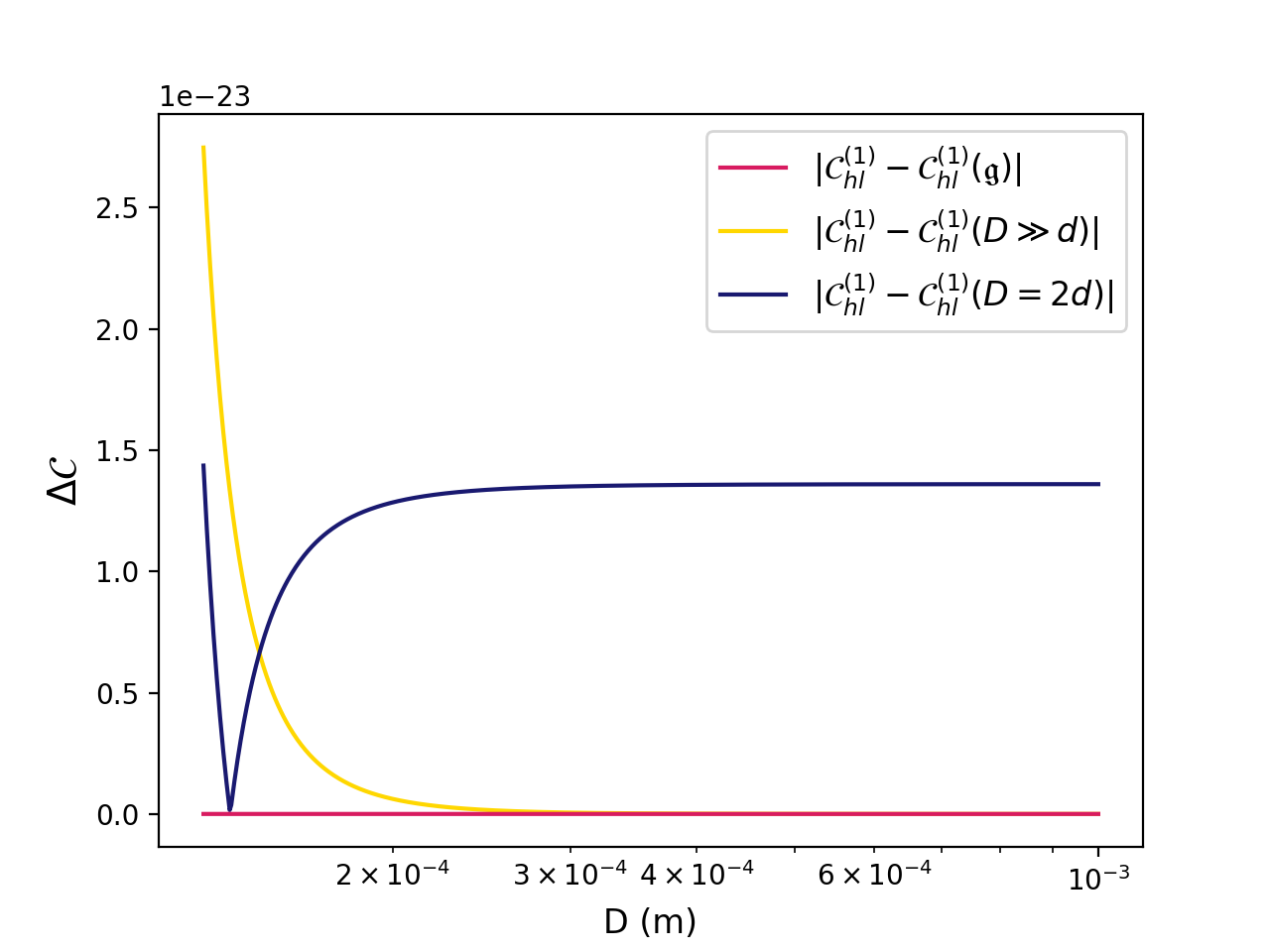}     
        \caption{Difference between the concurrence in Eq. \eqref{eq:conc_first_order} and the approximate concurrences in Eqs. \eqref{eq:concD}, \eqref{eq:concd} and \eqref{eq:concg} as a function of the distance $D$. For $d=10^{-4}$\,m, $m=10^{-14}$\,kg, $\omega_l=10^8$\,Hz, $\omega_h=10^8$\,Hz, $M=10^{-8}$\, kg.}
        \label{fig:approx_comp}
\end{figure}
These three limits work in different domains of $d/D$. 
In Fig. \ref{fig:approx_comp} we compare the different approximations as a function of $D$.
The range of $D$ shown is from $d$ (which is taken to be of the order $10^{-4}$\,m, following~\citep{Bose:2022uxe}) to $10^{-3}$\,m, the lines continue to be a constant for a larger $D$.
As one would expect, the concurrence $\mathcal{C}_{D=2d}$ is the worst approximation (except when $D=2d$).
The concurrence $\mathcal{C}_\mathfrak{g}$ performs the best across the whole range.
Although the concurrence $\mathcal{C}_{D\gg d}$ starts performing well around $D \sim 10^{-3}$\,m as well. 
We have explored and analysed these limits in order to be able to perform an analytical analysis in Sec. \ref{sec:results}.

The concurrence quantifies the entanglement due to the coupling between the light and heavy system.
Since entanglement and decoherence are two sides of the same coin, the concurrence between the subsystems provides a handle on the decoherence behaviour of the test masses due to the presence of the apparatus.
If there is no interaction between the heavy and light subsystems ($\mathfrak{g}_\pm =0$), then there is no gravitational decoherence from the experimental apparatuses. 
However, in any experiment the gravitational decoherence due to the experimental apparatus is unavoidable.
Minimizing the mass $M$, and maximizing the trap frequency $\omega$, as well as the distance $D$, minimizes the decoherence from the apparatuses.

\section{Contribution from higher order couplings}\label{sec:higher_order}

At first order the coupling between the heavy and light systems is only between the position operators (it is a quadratic coupling in the Hamiltonian, i.e, linear equations of motion). 
We now look at the post-Newtonian corrections which contains also momentum operators, focusing on cubic couplings in the Hamiltonian (quadratic couplings in the equations of motion)
Inserting the position operators in Eqs. \eqref{eq:pos_op_l} and \eqref{eq:pos_op_h} into the Hamiltonian given in Eq. \eqref{eq:H_g_op}, we obtain the cubic couplings in Eq. \eqref{eq:next_order_Hhl}, where we consider only the next order coupling between the light and heavy matter systems~\footnote{
The heavy-heavy and light-light couplings can be seen as self-interactions for the light-heavy bipartition used to calculate $\mathcal{C}_{\text{hl}}$.
}.
This expression contains the couplings between three operators: two light momentum/position operators and one heavy position operator, or two heavy position operators and one light position operator.
The relevant non-zero coefficients for the final wavefunction defined in Eq. \eqref{eq:coefficients} can be found the same way as before, by filling in the mode operators of Eq.\eqref{eq:mode_op} into the interaction Hamiltonian in Eq. \eqref{eq:H_g_op}.
We find the following nonzero terms:
\begin{alignat}{2}
    C_{0102} = C_{1020} &= \frac{\mathfrak{g}^{-}_{1}}{2\omega_h + \omega_l} \, , \\
    C_{0120} = C_{1002} &= -\frac{\mathfrak{g}^{+}_{1}}{2\omega_h + \omega_l} \, , \\
    C_{0201} = C_{2010} &= \frac{\mathfrak{g}^{-}_{3} - \mathfrak{g}^{-}_{2}}{\omega_h + 2\omega_l} \, , \\
    C_{0210} = C_{2001} &= \frac{\mathfrak{g}^{+}_{2} - \mathfrak{g}^{+}_{3}}{\omega_h + 2\omega_l} \, ,
\end{alignat}
with the six different couplings defined by:
\begin{align}
    \mathfrak{g}^{\pm}_{1} &= \frac{12 \sqrt{2} G}{\omega_h(D \pm d)^4} \sqrt{\frac{m \hbar}{\omega_l}} \, ,\\
    \mathfrak{g}^{\pm}_{2} &= \frac{12 \sqrt{2} G}{\omega_l(D \pm d)^4} \sqrt{\frac{M \hbar}{\omega_h}} \, , \\
    \mathfrak{g}^{\pm}_{3} &= \frac{3 G \omega_l}{\sqrt{2} c^2 (D \pm d)^2} \sqrt{\frac{M \hbar}{\omega_h}} \, .
\end{align}
The "$-$"-labelled couplings arise due to interactions between neighbouring heavy and light oscillators, while the "$+$"-labelled couplings arise due to maximally separated heavy and light oscillators.
Moreover, we underline the fact that the $\mathfrak{g}_3$ couplings represent the interaction of two momentum operators with a position operator, while the $\mathfrak{g}_1$ and $\mathfrak{g}_2$ couplings are attributable to the product of three position operators.

Recalling that $C_{0000} = 1$, the perturbed wavefunction up to first-order from Eq. \eqref{eq:psi_f} is given by: 
\begin{align}
    \ket{\psi_f} = \frac{1}{\sqrt{\mathcal{N}}} &\Big[ \ket{0000} +\frac{\mathfrak{g}^{-}_{1}}{2\omega_h + \omega_l} (\ket{0102} + \ket{1020}) \nonumber \\
    &- \frac{\mathfrak{g}^{+}_{1}}{2\omega_h + \omega_l} (\ket{0120} + \ket{1002}) \nonumber \\ &+\frac{\mathfrak{g}^{-}_{3} - \mathfrak{g}^{-}_{2}}{\omega_h + 2\omega_l} (\ket{0201} + \ket{2010}) \nonumber \\
    &+ \frac{\mathfrak{g}^{+}_{2} - \mathfrak{g}^{+}_{3}}{\omega_h + 2\omega_l} (\ket{0210} + \ket{2001}) \Big]\, , \label{eq:psi_f_full_no}
\end{align}
where the normalization constant is now given by $\mathcal{N} = 1 + 2\left[\frac{(\mathfrak{g}_{1}^{-})^2 + (\mathfrak{g}_{1}^{+})^2}{(2\omega_h + \omega_l)^2} + \frac{(\mathfrak{g}_{3}^{-} - \mathfrak{g}_{2}^{-})^2 + (\mathfrak{g}_{2}^{+} - \mathfrak{g}_{3}^{+})^2}{ (\omega_h + 2\omega_l)^2} \right]$.
The concurrence is calculated using its definition in Eq. \eqref{eq:concurrence_def_pure} and presented in Eq. \eqref{eq:conc_three_op} in the Appendix.
The expression simplifies based on the assumption that the characteristic couplings over the associated frequency is significantly smaller than one, i.e.,
\begin{align}
    \frac{\mathfrak{g}_{1}^{\pm}}{2\omega_h + \omega_l} \ll 1, \hspace{1cm} \frac{\mathfrak{g}_{2,3}^{\pm}}{\omega_h + 2\omega_l} \ll 1 \, . \label{eq:coupling_freq_lim}
\end{align}
In this regime the concurrence simplifies to:
\begin{align}
    \mathcal{C}_\text{hl}^{(2)} \approx 2 \sqrt{\frac{(\mathfrak{g}_1^{-})^2+(\mathfrak{g}_1^{+})^2}{(2\omega_h+\omega_l)^2}+\frac{(\mathfrak{g}_2^{-} - \mathfrak{g}_3^{-})^2 + (\mathfrak{g}_2^{+} - \mathfrak{g}_3^{+})^2 }{(\omega_h+2\omega_l)^2}}. \label{eq:conc_3op_w}
\end{align}

Again, the concurrence quantify the decoherence of the light oscillators due to the heavy oscillators.
From Eq. \eqref{eq:conc_three_op} we see that the concurrence decreases as the couplings $\mathfrak{g}_i$ are set to zero, with the concurrence being zero when there is no more coupling between the system and the environment, meaning that there is no loss of coherence in the light subsystem.

In order to get a better idea of the parameter dependence we explore the approximation where the couplings $\mathfrak{g}_{1,2,3}^-$ dominate the respective $\mathfrak{g}_{1,2,3}^+$ couplings.
In addition we use the fact that the coupling $\mathfrak{g}_3$ is suppressed by a factor $1/c^2$ (for typical values of the distances and trap frequencies), leaving us with the couplings $\mathfrak{g}_{1,2}^{-}$. 
The concurrence then simplifies to:
\begin{align}
    \mathcal{C}_\text{hl}^{(2)}(\mathfrak{g}) \approx \frac{24 G \sqrt{2\hbar}}{(D-d)^4 \omega_h\omega_l} \sqrt{\frac{m \omega_l}{(\omega_l+2\omega_h)^2} + \frac{M\omega_h}{(2\omega_l+\omega_h)^2}}.
    \label{eq:conc_threeop_g}
\end{align}
We see that the second order coupling contribution is suppressed by $\sqrt{\hbar}$, and has a inverse quartic dependence on the distance.

\begin{figure}[h]
    \centering
    \includegraphics[width=\linewidth]{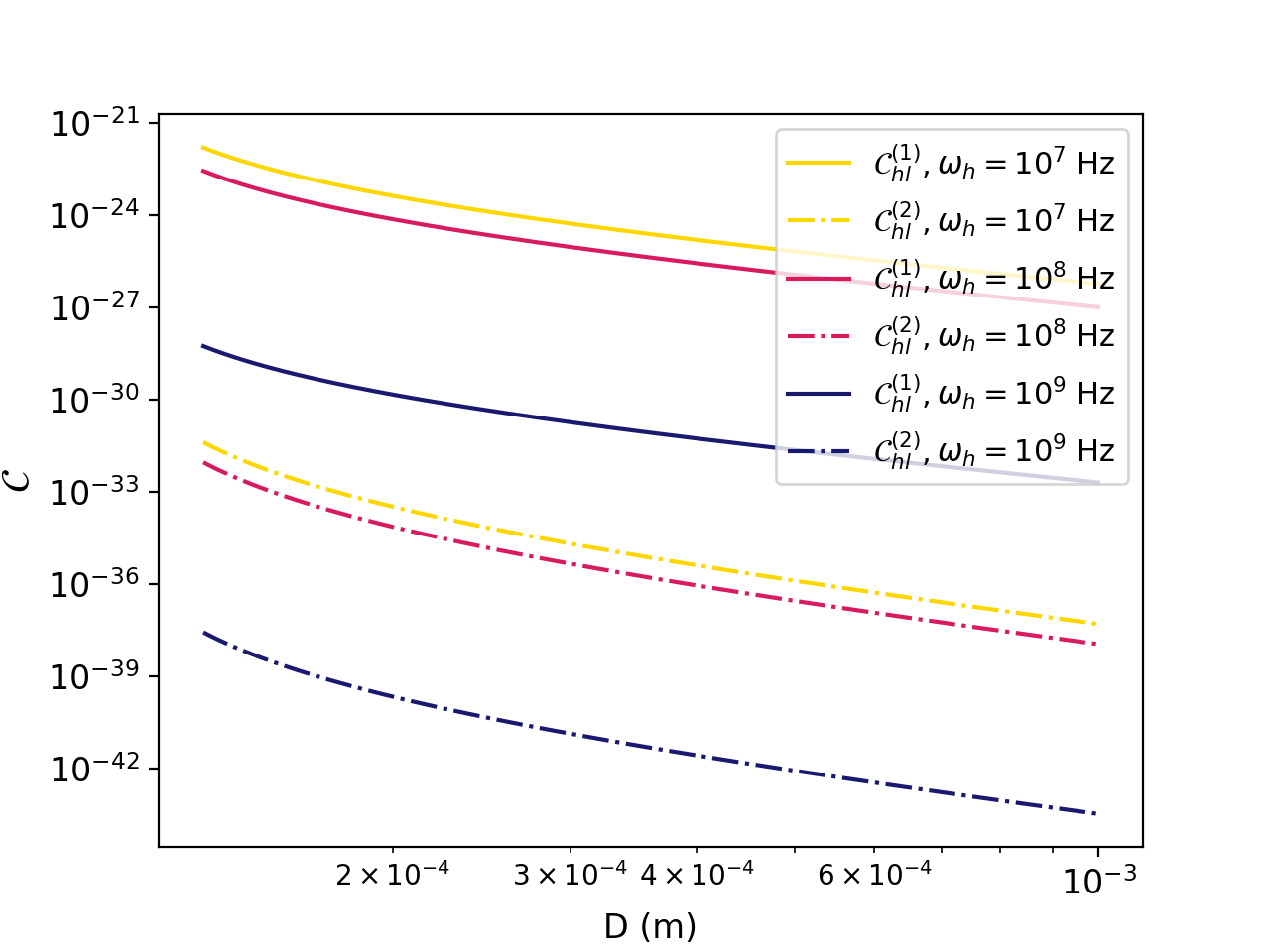}
    \caption{Concurrence as a function of the separation $D$. For $m\sim10^{-14}$\,kg, $d\sim10^{-4}$\,m and $\omega_l\sim10^{8}$\,Hz. For $M=10^{-8}$\,kg, and for different values of $\omega_h = 10^7, 10^8, 10^9$\,Hz. The solid lines represent the concurrence due to the first order couplings in Eq. \eqref{eq:concg}. The dashdotted represent the concurrence due to the next order couplings in Eq. \eqref{eq:conc_threeop_g}.}
    \label{fig:comparison}
\end{figure}

In Fig. \ref{fig:comparison} we plot the different order contributions to the concurrence given in Eqs. \eqref{eq:concg} and \eqref{eq:conc_threeop_g} for different $\omega_h$ and as a function of $D$.
The light oscillator system is taken to be as in Refs.~\citep{Bose:2017nin,Bose:2022uxe}.
The heavy frequencies are taken over a range $10^7-10^9$\,Hz, which are experimentally viable~\citep{Slezak:2018}.
The heavy mass is taken to be $~10^{-8}$\,kg, such that $M>m$.
We see that the first order concurrence dominates the next order concurrence with about ten orders of magnitude.
As $D$ increases the concurrence goes to zero and both order concurrences both becomes zero eventually.
This plot shows clearly that the next order coupling contributions to the decoherence are negligible. 

In this section we have calculated the decoherence due to next order momentum and position couplings of the system and environment. 
We saw that the dominant contribution comes from the coupling of the position operators, not the position-momentum operator coupling.
In Eq. \eqref{eq:lconc1} we saw that the momentum-contributions (at first order) also doesn't increase the light-light concurrence, $\mathcal{C}_\text{ll}$ much, they are suppressed by a factor $1/c^2$.
The contribution of the momentum terms in the decoherence scales as $\sqrt{\hbar}/c^2$, which is approximately an order of $1/c^2$ smaller.

Additionally we saw that these next order couplings entangle states where one of the light oscillators is in the first excited state and one of the heavy oscillators are in the second excited state.
This contribution is however dominated by the first order position couplings, which give rise to entanglement with first excited states.

\section{Restrictions on the experimental parameters}\label{sec:results}
\begin{figure}[t]
    \centering
    \includegraphics[width=\linewidth]{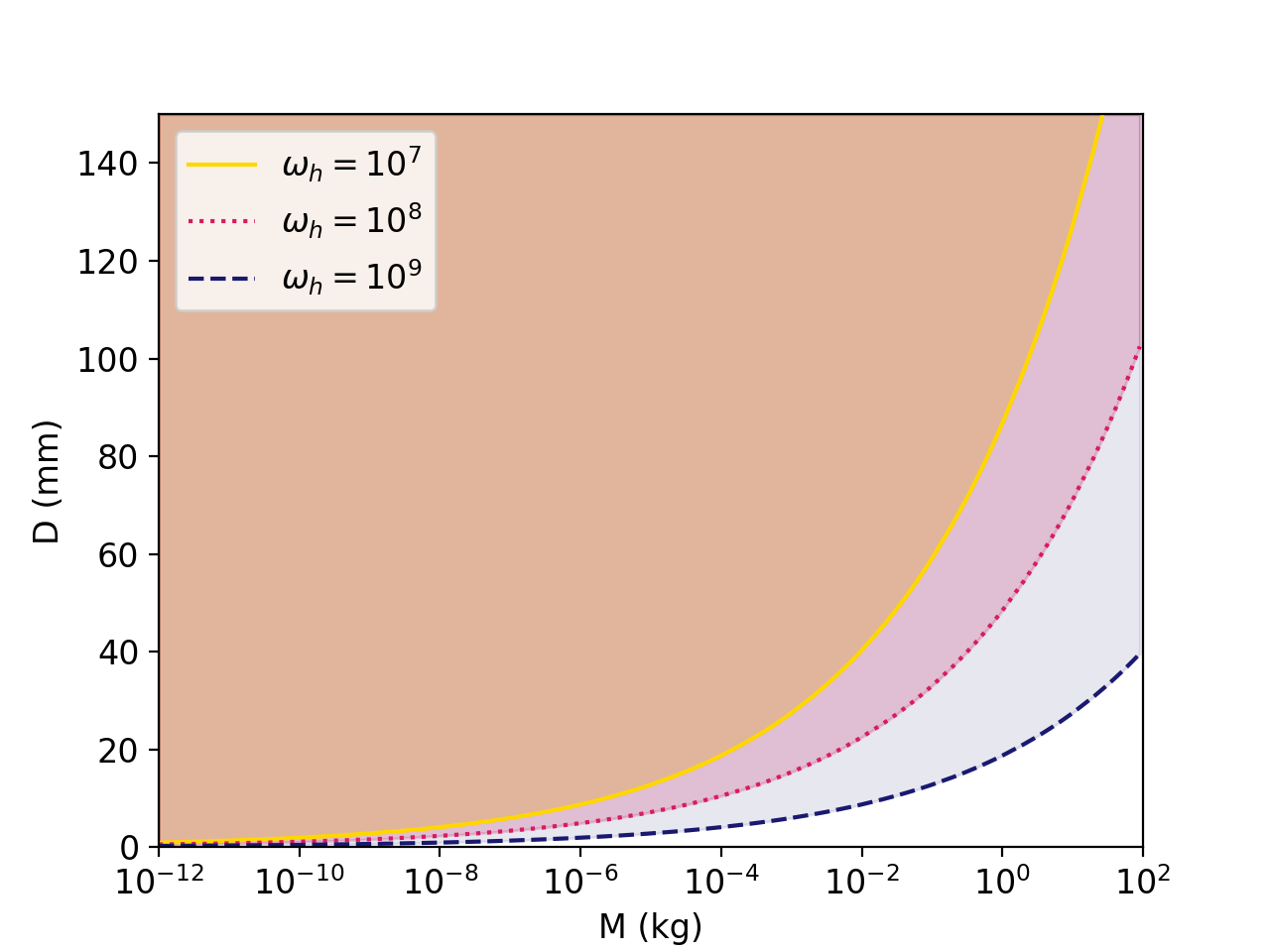}
    \caption{Distance $D$ as a function of the heavy oscillator mass $M$, given in Eq. \eqref{eq:ineq}, for different values of $\omega_h$.
    The shaded area above each curve indicates the parameter space such that the light-light concurrence dominates the decoherence.
    For $m\sim10^{-14}$\,kg, $d\sim10^{-4}$\,m and $\omega_l\sim10^{8}$\,Hz.
    }
    \label{fig:MvsD}
\end{figure}

In the above sections we found the decoherence from the heavy oscillators on the light oscillators.
We will now compare this decoherence to the concurrence between the two light test masses.
By requiring that the concurrence $\mathcal{C}_\text{ll} > \mathcal{C}_\text{hl}$, we aim to restrict the parameter space of the heavy system.
As we have seen that the momentum terms in $\mathcal{C}_\text{ll}$ and the second order couplings giving $\mathcal{C}_\text{hl}^{(2)}$ are heavily suppressed, we simply compare the $\mathcal{C}_\text{ll}$ and $\mathcal{C}_\text{hl}^{(1)}$ in the static case. 
So we require the first term in Eq. \eqref{eq:lconc1} to be larger than Eq. \eqref{eq:concg} (which uses the approximation that one of the coupling terms can be neglected, which was shown to be the best approximation across the range of $D$ considered).
The resulting inequality is:
\begin{equation}\label{eq:ineq}
    D > 
    \left( \frac{16 \sqrt{2M}\omega_l^2}{\sqrt{m \omega_l\omega_h}(\omega_l+\omega_h)} \right)^{1/3} d + d \, .
\end{equation}
This inequality is plotted in Fig. \ref{fig:MvsD}, where the light oscillator system parameter are chosen as found in previous works, $m\sim10^{-14}$\,kg, $d\sim10^{-4}$\,m and $\omega_l\sim10^{8}$\,Hz~\citep{Bose:2022uxe}.
In this figure the area above the curve is the parameter space such that the light-light entanglement dominates the decoherence.
The range of $M$ is chosen such that $M\gg m$.
We see that as $\omega_h$ increases, the allowed parameter space increases. 
Furthermore, a heavier apparatus mass requires a higher separation $D$ for the internal entanglement to dominate, as one would expect. 

The results derived from Fig. \ref{fig:MvsD} can be considered the results for the `static case', where the light oscillator system is considered to have no momentum.
We can also consider the case in which it does have momentum contributions, still at first order in the couplings.
This results in the inequality:
\begin{equation}\label{eq:ineq_mom}
    D > 
    \left( \frac{1}{\frac{1}{d^3\omega_l^2} + \frac{2}{d c^2}} \frac{16 \sqrt{2M}}{\sqrt{m \omega_l\omega_h}(\omega_l+\omega_h)} \right)^{1/3} + d \, ,
\end{equation}
which is similar to the one in Eq. \eqref{eq:ineq}. The second term in the denominator of the first fraction is the contribution from the momentum coupling in the light system. 
If this term is taken to be zero (so that it reduces to the static case), then we recover the Eq. \eqref{eq:ineq}. 
For the parameter space of the light system considered here ($d\sim10^{-4}$\,m, $\omega_l\sim10^8$\,Hz), the momentum contribution is of the order $10^{-12}$, and is thus negligibly small compared to the first term (which is of the order $10^{-4}$).
In this range of experimental parameters, the contribution to the entanglement from the momentum coupling within the light system is so heavily suppressed that it does not change the parameter space much.


The analysis we have done so far has compared the entanglement between the heavy and light system with the entanglement between the two light systems in the absence of the heavy system.
Comparing these two concurrences has provided a way to put restrictions on the parameter space. 
However, we should also have a look at the at the entanglement of the two light systems in the presence of the heavy systems.
By tracing out the heavy systems we can take the effects of the heavy system into account and then compute the concurrence within the light system.

\begin{figure}[t]
    \centering
    \includegraphics[width=\linewidth]{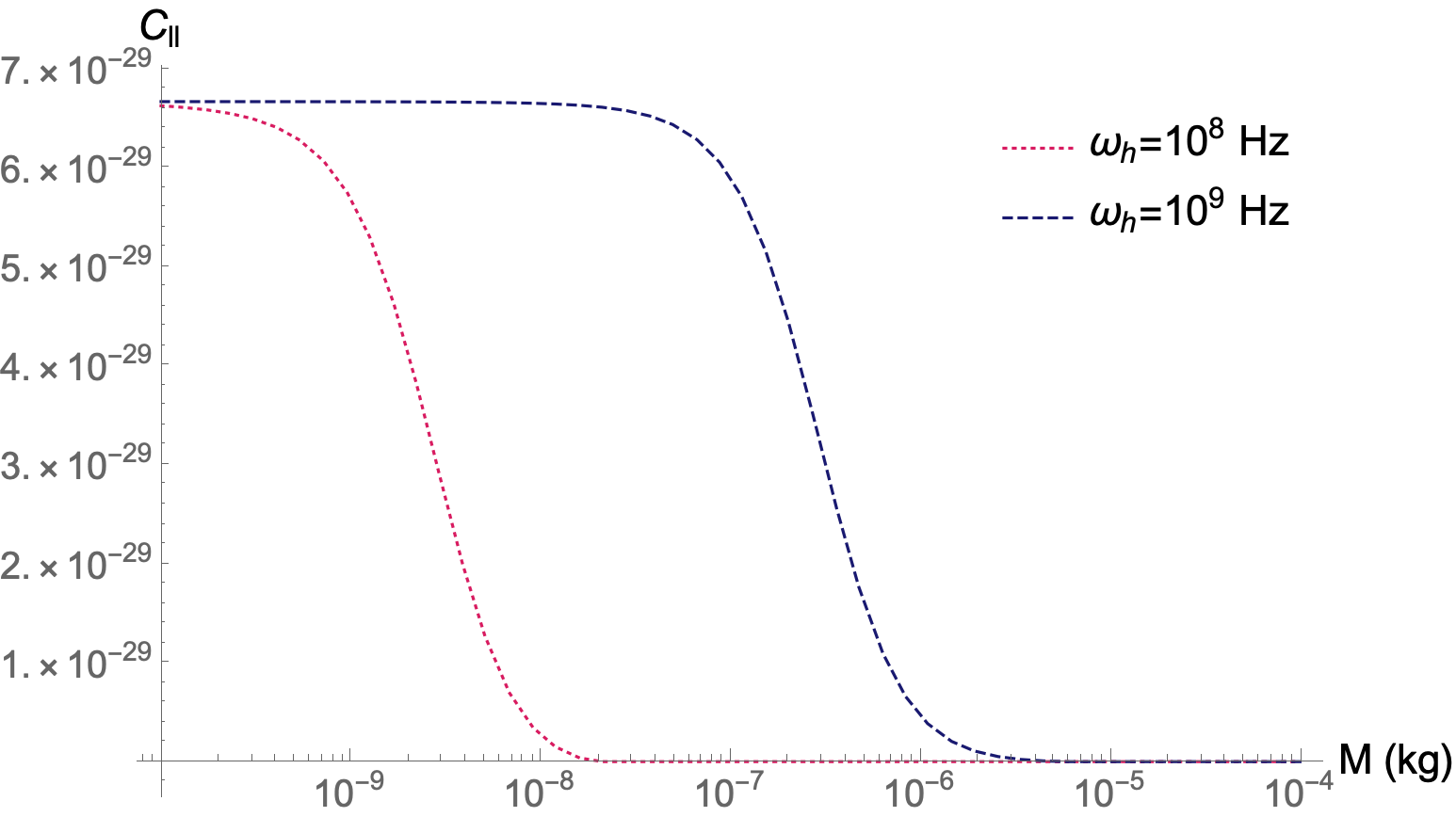}
    \caption{The concurrence between the two light oscillators in the presence of two heavy systems (resulting in decoherence), as a function of the heavy system's mass, $M$. 
    For light systems with parameters $m\sim10^{-14}$\,kg, $d\sim10^{-4}$\,m and $\omega_l\sim10^{8}$\,Hz. The heavy systems have $D=10^{-3}$\,m and $\omega_h=10^8-10^9$\,Hz.}
    \label{fig:conc}
\end{figure}

We consider the density matrix of the full system and we want to find the concurrence of the two light systems, given by the density matrix $\rho_1$ in Eq. \eqref{eq:density_matrix_ll}.
Since this represents a mixed state, we cannot use the definition of the concurrence given in Eq. \eqref{eq:concurrence_def_pure}.
Instead, we use the definition for the concurrence for mixed states:~\citep{Pathak:2013,Hill:1997pfa}
\begin{equation}\label{eq:concurrence_def_mixed}
    \mathcal{C} = \text{max}(0,\lambda_1-\lambda_2-\lambda_3-\lambda_4)\, ,
\end{equation}
where the $\lambda_i$'s are the ordered eigenvalues (highest to lowest) of the matrix $\sqrt{\sqrt{\rho_1}\Tilde{\rho}_1\sqrt{\rho_1}}$ with $\Tilde{\rho}_1 = (Y\otimes Y)\rho_1^*(Y\otimes Y)$, where $\rho^*_1$ is the complex conjugate of $\rho_1$, and $Y$ is the Pauli matrix~\footnote{
Previously we made the bipartition light-heavy, where the total $16\times16$ density matrix is pure. Therefore we were allowed to use the pure state definition of concurrence along this bipartition. Now we have traced out the heavy system, so for the bipartition light-light we have a total density matrix that is mixed, and we need to use the general (mixed state) definition. Note that this is different from taking a bipartition where system 1 contains one light system and system 2 contains the heavy systems and the other light system, which has a total pure state, and which regards the heavy systems as a part of the quantum system as opposed to the environment.
}.
In Fig. \ref{fig:conc} we plot the concurrence between the two light systems with the effect of the heavy systems taken into account, as a function of the heavy mass $M$. 
As expected, we see that as the heavy mass increases, the coupling between the heavy and light system increases and thus the entanglement between the two light systems decreases due to decoherence. 
As the heavy mass goes to zero, we recover the value for a static light-light system given by $\mathcal{C}_{ll}$ in Eq. \eqref{eq:lconc1}.
For the distance $D$ to be of the order of millimeters, the system fully decoheres if $M>10^{-8}$\,kg for $\omega_h=10^8$\,Hz and for $M>10^{-6}$\,kg for $\omega_h=10^9$\,Hz.
This is inline with the parameter space plotted in Fig. \ref{fig:MvsD} (at the order of millimeters, the lines have the mass values mentioned above). 
Additionally from this plot we could require that the decoherence reduces the entanglement to maximally $80\%$ of the original value, which would require the heavy mass to be approximately of order $10^{-9}$\, kg or smaller for $\omega_h=10^8$\,Hz.
Knowing the experimental parameters of the heavy system can provide us with information about the expected coherence of the light system.

\section{Discussion}\label{sec:conclusion}
In this paper we have investigated the gravitational decoherence induced by the experimental apparatus in the QGEM scheme. 
We have modelled the scheme with coupled harmonic oscillators: two light oscillators coupled to two heavy oscillators, the former (latter) two playing the role of the system (experimental apparatus).
Considering the apparatuses to be heavy, static oscillators and the test masses to be light non-static oscillators, we found the decoherence due the apparatuses on the test masses.
The decoherence results are given as concurrences in Eqs. \eqref{eq:conc_first_order} and \eqref{eq:conc_3op_w}.



We computed the concurrence for the quadratic and the cubic couplings, showing the dominant terms and their dependencies on the experimental parameters.
The first order coupling concurrence was found to be very dominant over the higher order contributions.
A large separation between the test masses and the apparatuses, high trap frequencies, and low masses of the apparatuses will reduce the decoherence, as expected. 
We explored the limits $D\gg d$ and $D=2d$ corresponding to different setups, resulting in the same dependence on the experimental parameters, but resulting in a bigger decoherence for the $D=2d$ setup due to the smaller $D$.
We also approximated the concurrence by assuming that the the nearest neighbour coupling dominates, which turned out to be the best approximation, and we used this  to  restrict the parameter space for the apparatus.

We also explored the first order momentum contributions to the decoherence, which appeared in the next order couplings and are therefore suppressed by a factor $\sqrt{\hbar}$ compared to the momentum contributions to the light-light entanglement, which entered at the lowest order couplings.
We found that relative to the static contributions to the entanglement, the momentum contributions are negligible 

By requiring the decoherence to be smaller than the light-light concurrence, we found that the separation $D$ will be of the order of centimetres for the masses upto $M\sim 100$\,kg if the trap frequency is larger than $10^8$\,Hz.
A smaller trap frequency for the same range of masses requires a larger separation. 
Of course, a larger separation $D$, a smaller mass $M$, and a higher frequency $\omega_h$, decrease the decoherence.
This is illustrated in Fig. \ref{fig:conc} in which we plotted the light-light entanglement under the influence of interactions with the environment (i.e., the heavy system).

By modelling the the apparatuses as harmonic oscillators, we are able to make an approximate prediction about the allowed separation between the detectors and the test masses that does not completely destroy the coherence of the test particles.
For example, the typical spacing of ion traps are of the order of millimetres, which is smaller then the scale found here, and the decoherence is smaller than the light-light entanglement only for masses $M$ upto $10^{-6}$\,kg, for the considered frequencies (as seen from Fig. \ref{fig:MvsD}). 

Setting one of the heavy masses to be zero, $M_B=0$, we can also use our method to find the decoherence due to a single massive oscillator. At no point in the calculations have we assumed that $M>m$, therefore the resulting decoherence rates hold for any mass $M$. 
However, in the range where $M<m$, we expect the light-light entanglement to be dominant since the gravitational coupling scales with the masses, assuming that the distances are such that $D>d$.
In other words, these light sources of decoherence might become relevant at very short distances.
Similarly we have not explored masses of $M\sim m$, where the coupling between the heavy and light system is of the same strength. 
These sources are expected to become relevant at $D\sim d$.
Our results for the decoherence rate are general and can also be used for other mass ranges.
We have modelled the mass $M$ as a coherent state, and future research could also explore different type of states, such as thermal states.

\section*{Acknowledgements} \label{sec:acknowledgements}
MT would like to acknowledge funding by the Leverhulme Trust (RPG-2020-197). 
MS is supported by the Fundamentals of the Universe research program within the University of Groningen. 
AM’s research is funded by the Netherlands Organisation for Science and Research (NWO) grant number 680-91-119.



\onecolumngrid
\appendix
\section{Equations}
Interaction Hamiltonian between the heavy and light system, after a non-relativistic expansion upto $\order{1/c^4}$ and at first order in $G$:
\begin{align}
\Delta\hat{H}_\text{g} &=
    -G \bigg[ \frac{m M}{\abs{\hat{x}_a-\hat{x}_A} }+\frac{m M}{\abs{\hat{x}_a-\hat{x}_B} }+\frac{m M}{\abs{ \hat{x}_A-\hat{x}_b} }+\frac{m M}{\abs{ \hat{x}_b-\hat{x}_B} }+\frac{m^2}{\abs{ \hat{x}_a-\hat{x}_b} }+\frac{M^2}{\abs{ \hat{x}_A-\hat{x}_B} }\bigg] \nonumber \\
    &-\frac{G}{c^2} \bigg[ \frac{3 M}{2 m} \bigg(\frac{\hat{p}_a^2}{\abs{ \hat{x}_a-\hat{x}_A} }+\frac{\hat{p}_a^2}{\abs{ \hat{x}_a-\hat{x}_B} }+\frac{\hat{p}_b^2}{\abs{ \hat{x}_A-\hat{x}_b} }+\frac{\hat{p}_b^2}{\abs{ \hat{x}_b-\hat{x}_B} }\bigg)+\frac{3 \hat{p}_a^2-8 \hat{p}_a \hat{p}_b+3 \hat{p}_b^2}{2 \abs{ \hat{x}_a-\hat{x}_b} } \bigg] \nonumber \\
    &-\frac{G}{c^4} \bigg[ \frac{5M}{8m^3} \bigg( \frac{\hat{p}_a^4}{\abs{ \hat{x}_a-\hat{x}_A} }+\frac{\hat{p}_a^4}{\abs{ \hat{x}_a-\hat{x}_B} }+ \frac{\hat{p}_b^4}{\abs{ \hat{x}_A-\hat{x}_b} }+\frac{ \hat{p}_b^4}{\abs{ \hat{x}_b-\hat{x}_B} } \bigg)
    + \frac{5 \hat{p}_a^4 -18 \hat{p}_a^2 \hat{p}_b^2+5 \hat{p}_b^4}{8 m^2 \abs{ \hat{x}_a-\hat{x}_b} }\bigg]+\order{\frac{1}{c^6}} \, . \label{eq:H_g_op}
\end{align}

Interaction Hamiltonian between the heavy and light system in terms of the position operators $\delta\hat{x}_{a,b} = \hat{x}_{a,b} \pm d/2$ and $\delta\hat{x}_{A,B} = \hat{x}_{A,B} \pm D/2$, only showing terms containing three operators (which are the second order interactions, as indicated by the superscript $(2)$, giving rise to quadratic terms in the equations of motion):
\begin{align}
    \hat{H}_{hl}^{(2)} 
    &= 48 G m M \bigg[\frac{\delta\hat{x}_a (\delta\hat{x}_A)^2-(\delta\hat{x}_a)^2 \delta\hat{x}_A}{(D-d)^4} +\frac{\delta\hat{x}_a (\delta\hat{x}_B)^2-(\delta\hat{x}_a)^2 \delta\hat{x}_B}{(d+D)^4} \nonumber \\
    &\hspace{2cm}+\frac{\delta\hat{x}_A (\delta\hat{x}_b)^2-(\delta\hat{x}_A)^2 \delta\hat{x}_b}{(d+D)^4}+\frac{\delta\hat{x}_b (\delta\hat{x}_B)^2-(\delta\hat{x}_b)^2 \delta\hat{x}_B}{(D-d)^4}\bigg] \nonumber \\
    &+ \frac{6 G M}{c^2 m} \left[\frac{(\hat{p}_b)^2 \delta\hat{x}_A + (\hat{p}_a)^2 \delta\hat{x}_B}{(d+D)^2} - \frac{(\hat{p}_a)^2 \delta\hat{x}_A + (\hat{p}_b)^2 \delta\hat{x}_B}{(D-d)^2}\right] \, . \label{eq:next_order_Hhl}
\end{align}


Concurrence between the heavy and light system, arising from the interaction in Eq. \eqref{eq:next_order_Hhl} above:
\begin{align}
    \mathcal{C}^{(2)}_\text{hl} &= \Bigg\{2 - \frac{2}{\mathcal{N}^2} \bigg[1 + \frac{2 \left[( \mathfrak{g}_{3}^{-} - \mathfrak{g}_{2}^{-} )^2 + ( \mathfrak{g}_{2}^{+} - \mathfrak{g}_{3}^{+} )^2 \right]^2 + 8 \left(\mathfrak{g}_{3}^{-} - \mathfrak{g}_{2}^{-} \right)^2 \left(\mathfrak{g}_{2}^{+} - \mathfrak{g}_{3}^{+} \right)^2 }{\left(\omega_h + 2 \omega_l\right)^4 } \nonumber \\ & \hspace{2.5cm}+ \frac{2 \left[( \mathfrak{g}_{1}^{+} )^2 + (\mathfrak{g}_{1}^{-} )^2 \right]^2 + 8 (\mathfrak{g}_{1}^{+} \mathfrak{g}_{1}^{-} )^2 }{\left(2 \omega_h + \omega_l\right)^4} \bigg] \Bigg\}^{1/2} \, .\label{eq:conc_three_op}
\end{align}

\end{document}